# Stability Analysis of C-band 500-kW Klystron with Multi-cell Output cavity

**Jihyun Hwang**

*Department of Physics, POSTECH, Pohang 37673*

**Sung-Ju Park and Won Namkung**

*Pohang Accelerator Laboratory, Pohang 37874*

**Moohyun Cho**

*Department of Advanced Nuclear Engineering, POSTECH, Pohang 37673*

A prototype 5-GHz 500-kW CW klystron (model E3762 provided by Toshiba Electron Tubes & Devices Co. Ltd.) has been operating as the RF source for the lower hybrid current drive (LHCD) system in the KSTAR tokamak. In order to investigate how the efficiency of the 5-GHz 500-kW CW klystron prototype could be enhanced, the cavity design study is being carried out with simulation code based on the main klystron operation parameters. This is being done by simulating klystron performances for various cavity parameters including the number of cavities, inter-cavity distance, and cavity tuning frequencies. The simulation has been done with the FCI (field charge interaction) code aided by a matlab script for scanning input parameters. Initial set of scan parameters was obtained by benchmarking the E3762 klystron. It was possible to obtain optimized design parameters with better efficiency for a cavity system adopting a multi-cell output cavity. However, the multi-cell output cavity is prone to produce a self-oscillation due to the prolonged (several half RF periods) beam-field interaction along its multiple gaps. We have checked the feasibility of the optimization by evaluating the stability of output cavity system. The stability is given by the ratio of a beam-loading conductance to the circuit conductance.

Email: jhyun@postech.ac.kr

Fax: +82-54-279-2758

# I. INTRODUCTION

Klystron [1] is a vacuum electron device used for amplifying RF signals in the wavelength range from several meters to sub-millimeters. Klystrons consist of an electron gun to produce electron beams, a solenoid magnet or PPM (periodic permanent magnets) to focus the electron beam, cavity section to couple, amplify and extract RF signals by interacting with the electron beam, and a collector to dump the spent electron beams at the end. The cavity section is important because it determines klystron performances including efficiency, gain and bandwidth. Design of the cavity section includes determining the number of cavities, cavity tuning frequencies and inter-cavity distances.

An axial extension of the interaction region by use of the multi-gap cavity [2-10] is suggested by Chodorow [3] and Wessel-Berg [4-5] in the early 1960s. They analyzed the behavior of the multi-gap cavity with extended interaction fields based on space charge theory. Preist and Leidigh [6-7] discussed the stability of the extended interaction cavity which is a major limitation of the improvement in performance and the behavior of the stability parameter of the tube is dominated by variations in the loaded quality factor $Q_L$. Lien and Robinson [8-10] analyzed the stability of a double-gap coupled output cavity based on an equivalent circuit. The multi-gap cavity has not only higher $R/Q$, which leads to a wider bandwidth and a higher circuit efficiency relative to the conventional klystron, but also a large surface reducing the RF gradient and the danger of RF breakdown. On the other hand, the multi-gap cavity can acts as an oscillator, called a self-oscillation, resulting in a low output power, when the magnitude of beam-loading conductance is larger than the magnitude of circuit conductance.

The prototype 500-kW CW klystron operating at 5 GHz is used as a RF source in the LHCD system of the KSTAR tokamak. The design parameter and target performance are shown in Table 1. We optimized the cavity system of 500-kW CW C-band klystron including the multi-cell output cavity which is used to reduce a heat load of each cell and prevent RF breakdowns. The multi-cell cavity can provide the higher power and the wider bandwidth, however there is a possibility to generate the self-oscillation which leads to decreased output power. In order to avoid this, we analyze the stability for the designed output cavity system.

## II. BENCHMARKING OF EXISTING DESIGN

In order to obtaining a set of starting parameters for our optimization, we benchmark the E3762 of Toshiba Electron Tubes and Devices Co. Ltd. (TETD) as a starting parameter. There are six cavities which consist of one input cavity, one multi-cell output cavity and four intermediate cavities including a second harmonic cavity. The inter-cavity distances are shown in Fig. 1.

White-noise method [11-12] has been used to measure the resonant frequencies of cavities in multi-cavity klystrons. In this method, a weak electron beam is used as a probe to measure the tuning of cavity. As shown Fig. 2, an input cavity of klystron is connected to a RF amplifier and a sweep signal generator which provides sinusoidal signal with the frequency range of 4 to 6 GHz. Signal from the output cavity is measured by a spectrum analyzer attached to the output waveguide. For this measurement, the cathode voltage is reduced to 3 kV to generate the weak electron beam. Current to the bucking coil was also carefully adjusted to achieve optimum beam profile which results in clear signals in the spectrum analyzer. When the electron beam pass through the cavities, beam-cavity interaction occurs exchanging energy between the beam and cavity. Since the beam-cavity interactions are strong around the resonant frequencies of cavities, information on the cavity resonant frequencies is contained within the electron beam and coupled out at the output cavity. Finally the resonant frequencies of cavities appear as peaks or dips at the spectrum analyzer. The results of the white-noise measurement for the E3762 klystron are listed in Table 2. While three of four intermediate cavities were clearly detected, the third cavity (second harmonic cavity) was not. The resonant frequency of the second harmonic cavity is believed to be around 10 GHz.

We also measured focusing magnetic field, required to control the electron beam along the interaction region, for the E3762. During the measurement excitation current to its solenoid coils is decreased to 1/10 of normal operation value due to a cooling problem. This magnetic field was measured at six points along solenoid axis. Using the POISSON code, we determined solenoid magnet geometry which fits to the measured profile. Magnetic field profile (as shown in Fig. 3) obtained from the POISSON simulation for the determined solenoid geometry was used in later PIC simulations.

### III. PIC SIMULATION

The field charge interaction code FCI [13-14] is a 2+1/2 dimensional particle-in-cell simulation code, which has been used to analyze the beam dynamics of klystrons. The FCI code simulates beam motion in a drift-tube section by taking into account the space charge fields, the self-magnetic field as well as the external focusing field.

In the FCI code, the beam is typically axially symmetric in shape and assumed originating from a Pierce gun and focused by a solenoid magnetic field. The beam is represented by the beam voltage, current, beam radius and its slope at the entrance to the problem boundary. The transverse profile is greatly simplified and the current values (relative to the beam axis) at only the beam edge and half edge are specified. Unlike general PIC codes, cavities are modeled by equivalent circuits with their gaps on the drift tube wall. The spatial distribution of the gap field, which is required for calculating the beam-cavity interaction, is computed by an external field solver, DENKAI. The equivalent circuit constants are derived from the cold cavity parameters, the resonant frequency $f_0$, the shunt impedance $R/Q$ and loaded quality factor $Q_L$. In the case of a multi-cell coupled cavity, coupling between cells is represented by introducing the coupling constant $k_{12} = \frac{M_{12}}{\sqrt{L_1 L_2}}$ where $M_{12}$ is the mutual inductance. $L_1$ and $L_2$ are inductances of two cells which are coupled each other. RF field of the multi-cell output cavity is represented by the superposition of cell modes of which the amplitudes and phases as well as the spatial distributions of the gap fields are independently specified.

The resonant frequencies of intermediate cavities except the third cavity are measured by white-noise method as Table 2. The optimum value of resonant frequency for the third cavity was found 9955 MHz from FCI simulations. A three-cell cavity is used as the output cavity, so three resonant frequencies, $f_1$, $f_2$ and $f_3$, and the coupling constants, $k_{12}$, $k_{23}$ and $k_{13}$, have to be determined. To find out these unknown values, the frequency of each cell was scanned from 4700 to 5300 MHz with different sets of coupling constants such as $k_{12} = 0.1$, $k_{23} = 0.1$ and $k_{13} = 0.01$ or $k_{12} = 0.2$, $k_{23} = 0.2$ and $k_{13} = 0.02$. The other parameters for this simulation are listed in Table 3. Fig. 4 and Fig. 5 shows the output power and cell dissipation powers with varying cell tuning frequencies. The cell tuning frequencies ($f_1$, $f_2$ and $f_3$ ) range from 4700 to 5300 MHz with step of 50 MHz. In the Fig. 4 and Fig. 5, index means to count out the number of scan steps in $f_1$, $f_2$ and $f_3$. One of the design requirements of the output cavity system is the output power greater than 500 kW which should be achieved at the cell dissipation power less than 5 kW. The exact value of the cell dissipation limit has to be determined in coming work. The gap voltage of each cell should be minimized to prevent RF breakdowns. While these requirements were met with the coupling constant of 0.2, coupling constant less than 0.1 was not practical even though the output power was greater than 500 kW. This was because the RF gradient and dissipation power were too high as shown in Fig. 4(a) and Fig. 5(a). Coupling constant of 0.3 was also impractical because the RF field of the first cell was too high as shown in Fig. 4(c) and Fig. 5(c). Table 4 shows sets of output cavity parameters which were found to satisfy the design requirement.

## IV. DEFINITION OF STABILITY

The total power loss of the cavity $P_T$, the sum of ohmic loss $P_0$, the transmitted power to an external circuit $P_{ext}$ and the power transferred to the electron beam $P_b$, must be positive to prevent the self-oscillation [4-5, 15]:

$$P_T = P_0 + P_{ext} + P_b > 0 . \tag{1}$$

The stability is generally expressed in quality factor, indicating the energy flow in the cavity system. The reciprocal of the total quality factor is defined as

$$\frac{1}{Q_T} = \frac{P_T}{\omega W_s} = \frac{(P_0 + P_{ext} + P_b)}{\omega W_s} \tag{2}$$

where $W_s$ is the stored energy and $P_T$ is the total power loss. So, (1) can be expressed in the form of the quality factor:

$$\frac{1}{Q_T} = \frac{1}{Q_0} + \frac{1}{Q_{ext}} + \frac{1}{Q_b} = \frac{1}{Q_L} + \frac{1}{Q_b} > 0 . \tag{3}$$

The reciprocal of the beam-loading quality factor, the quality factor with the beam present, is given by

$$\frac{1}{Q_b} = G_b \left(\frac{R}{Q}\right) \tag{4}$$

where $G_b$ is the beam-loading conductance, which represents energy transfer between the beam and the cavity. It can be either a positive or negative number depending on the direction of power flow between the beam and the cavity. The self-oscillation is caused by the surplus energy, accumulated when the power flowed to the cavity overwhelms the sum of ohmic loss and external loss. .

The formula of the normalized beam-loading conductance based on the space charge wave theory [6] for small signals is given by

$$\frac{G_b}{G_0} = \frac{1}{8} \frac{\beta_e}{\beta_q} \left[M^2(\beta_e - \beta_q) - M^2(\beta_e + \beta_q)\right] \tag{5}$$

where $G_0 = I_0/V_0$ is the beam conductance, $\beta_e = \omega/u_0$ is the electronic propagation constant and $\beta_q = \omega_q/u_0$ is the reduced plasma propagation constant with the velocity of the electron beam $u_0$ and the reduced plasma frequency $\omega_q$. The $M(\beta_e - \beta_q)$ and $M(\beta_e + \beta_q)$ are the coupling coefficient of a fast and slow space charge wave which is written by

$$M(\beta_e \pm \beta_q) = \frac{\int E_z(z) e^{-i(\beta_e \pm \beta_q)z} dz}{\int |E_z(z)| dz} \tag{6}$$

where $E_z(z)$ is an axial electric field of the cavity.

The stability of the system [7-8, 16] is expressed as

$$S = \frac{1/Q_t - 1/Q_L}{1/Q_L} = \frac{1/Q_b}{1/Q_L} = \frac{Q_L}{Q_b} \tag{7}$$

$$\frac{1}{Q_L} = \frac{1}{Q_0} + \frac{1}{Q_{ext}} \tag{8}$$

where $Q_L$ is the loaded quality factor without the beam. The condition with $S > 0$ corresponds to the positive beam-loading conductance which means the beam absorb energy from the cavity. Therefore, the system is stable regardless of the external load $Q_{ext}$. $-1 < S < 0$, i.e., $Q_L < |Q_b|$ implies the negative beam-loading conductance, the energy transfer from the beam to the cavity. The system is stable when the external load and the cavity can fully absorb the energy released by the beam. In the case of the multi-cell output cavity which extracts the energy to the external load, the beam-loading conductance must be negative, but for $S < -1$, i.e., $Q_L > |Q_b|$, the system is unstable.

## V. STABILITY ANALYSIS

The normalized beam-loading conductance $G_b/G_0$ depends on the coupling coefficient, the interaction between the beam and the cavity, and the relationship between the beam velocity and the axial velocity of the RF wave as (5). The coupling coefficient M is a function of the axial electric field. We utilized the DENKAI code [17] to obtain axial electric fields of each cell (the cell fields). The electric field distribution of three-cell coupled cavity can be obtained by adding the cell fields. The cell phases and coupling constants between cells are also considered in equivalent circuit model for a coupled cavity. Fig. 6 shows the simulated electric field patterns for three normal modes. Using these axial electric field, the coupling coefficient and the normalized beam-loading conductance are calculated.

The normalized beam-loading conductance is computed by varying the beam voltage from 5 to 100 kV and the phase shift from 0 to 360 degree as shown Fig. 7. When the beam-loading conductance is negative at around 68 kV, the operating beam voltage, the values of phase shift is proper operating mode. In the same manner, Fig. 8(a) and (b) show the results of changing the phase shift from cell to cell and the distance between cells. The beam-loading conductance is negative when the transit angle $\beta_e l_{12}$ is from 3.5 to 5 radian at the region of phase shift, resulting in the negative value in Fig. 7. The distance $l_{12}$ between first and second cell is about from 15 to 26 mm with the electron propagating constant $\beta_e = \omega/u_0$ at 68 kV. Likewise, the transit angle $\beta_e l_{23}$ should be from 3 to 5 radian of which the distance $l_{23}$ is about from 13 to 22 mm.

To check the stability of output cavity system designed by the FCI, the cell phases resulting from the cell frequencies scanned (from 4700 to 5300 MHz) in the FCI must be computed. The cell phase is calculated from the cavity impedance which is given by

$$Z = \frac{Z_0}{1+j2Q_L\delta} \quad where \; \delta = \frac{f_0-f_c}{f_c} \tag{9}$$

$$\text{phase} = \text{atan}\left(\frac{Im(Z)}{Re(Z)}\right) \cong \text{atan}(2Q_L\delta) \tag{10}$$

Using the phase computed by (10) and the coupling constants, the electric field distribution and the beam-loading conductance can be obtained. Fig. 9 shows the electric field of designed output cavity. Both modes shown in Fig. 9(a) and (b), $3\pi/2$- and $2\pi/3$-mode like field distributions respectively, well correspond to the region of the negative beam-loading conductance of Fig. 7. The distance between cells in Fig. 1 also corresponds to the computed distances of Fig 8(a) and (b).

In Fig. 10, we compare the output power (simulated by the FCI) and the normalized beam-loading conductance with varying cell frequencies for different sets of coupling constant and a fixed beam voltage 68 kV. The region of the negative beam-loading conductance reasonably matches that of high output power.

The reciprocal loaded quality factor without the beam of the three-cell output cavity is derived as

$$\frac{1}{Q_L} = \left(\frac{3}{Q_0} + \frac{1}{Q_e}\right) \cong \frac{1}{Q_e} \quad where \; Q_0 \gg Q_e \; . \tag{11}$$

In the case of the output cavity, the beam-loading quality factor $Q_b$ must be negative and its absolute value be larger than the loaded quality factor $Q_L$ for the stable system. According to Table 3 and (11), the value of the loaded quality factor is 25, thus the magnitude of reciprocal beam-loading quality factor must be smaller than ~0.04. On the other hand, the required beam-loading quality factor for the stable system can be changed according to the external coupling, related to the value of loaded quality factor. As shown Fig. 11, the reciprocal beam-loading conductance of the designed output cavity system is smaller than ~0.04, so the system is stable.

## VI. CONCLUSION

Based on the white-noise measurement of existing 5 GHz klystron (TETD E3762), the cavity system is re-designed by using the FCI and DENKAI codes and investigated its stability. The three-cell output cavity is being fully optimized by scanning various parameters including frequencies of each cell and coupling constants between adjacent cells. Among these, there are few sets of frequencies with a certain set of coupling constants satisfying the requirements of output power and RF gradient for each cell. In the light of formulism of the beam-loading conductance, the stability of the designed output cavity system is examined. We found that the magnitude of computed reciprocal beam-loading quality factor was smaller than that of the reciprocal loaded quality factor and therefore the designed system was stable.

## ACKNOWLEDGEMENT

This work was supported by National R&D Program (grant number: 2014M1A7A1A02029891, 2016R1A6B2A01016828), BK21+ program through the National Research Foundation of Korea (NRF) and the ITER technology R&D program funded by the Ministry of Science, ICT and Future Planning, Korea.

Table 1. Design parameter and target performance.

| Parameter | Target design |
|---|---|
| Beam voltage | 68 kV |
| Beam current | 15 A |
| Drive frequency | 5 GHz |
| Drive power | Max. 30 W |
| Peak output power | 500 kW |
| Efficiency | > 50 % |
| Saturated gain | > 42 dB |
| μ-perveance | 0.85 A/V$^{3/2}$ |

Table 2. Resonant frequency measured by white-noise method: The third cavity is 2$^{nd}$ harmonic cavity of which the resonant frequency is believed to be near 10 GHz. There is no meaningful results due to too much noise.

| No. of Cavity | Resonant frequency [MHz] |
|---|---|
| 2 | 4997 |
| 3 | No result |
| 4 | 5077 |
| 5 | 5092 |

Table 3. Simulation condition.

| Parameter | Target design |
|---|---|
| Beam voltage | 68 kV |
| Beam current | 15 A |
| Drift tube radius | 0.4 cm |
| Beam radius | 0.32 cm |
| Drive power | 5 W |
| $Q_L$ of input cavity | 363 |
| $Q_L$ of output cavity | 25 |
| R/Q of 1$^{st}$ cell | 80.86 |

| R/Q of 2$^{nd}$ cell | 77.45 |
|---|---|
| R/Q of 3$^{rd}$ cell | 81.78 |

Table 4. Sets of resonant frequency of the multi-cell output cavity satisfies the requirement.

(a) $k_{12} = 0.2$, $k_{23} = 0.2$ and $k_{13} = 0.02$

| $f_1$ [MHz] | $f_2$ [MHz] | $f_3$ [MHz] | $P_{out}$ [kW] | $P_d$ of 1$^{st}$ cell [W] | $P_d$ of 2$^{nd}$ cell [W] |
|---|---|---|---|---|---|
| 4700 | 5000 | 5250 | 566 | 4150 | 4510 |
| 4700 | 5050 | 5200 | 570 | 3160 | 3440 |
| 4700 | 5100 | 5200 | 525 | 2280 | 3030 |

(b) $k_{12} = 0.2$, $k_{23} = 0.25$ and $k_{13} = 0.02$

| $f_1$ [MHz] | $f_2$ [MHz] | $f_3$ [MHz] | $P_{out}$ [kW] | $P_d$ of 1$^{st}$ cell [W] | $P_d$ of 2$^{nd}$ cell [W] |
|---|---|---|---|---|---|
| 4700 | 5100 | 5300 | 590 | 3450 | 3960 |
| 4700 | 5150 | 5250 | 574 | 2720 | 3080 |
| 4750 | 5100 | 5250 | 557 | 3760 | 2810 |

Figure Captions.

Fig. 1. Schematic drawing of C-band klystron. Unit is in cm.

Fig. 2. Experiment setup of the white-noise method.

Fig. 3. Focusing field profile as a function of the axial distance.

Fig. 4. Output power of the scanned simulation with

(a) $k_{12} = 0.1$, $k_{23} = 0.1$ and $k_{13} = 0.01$.

(b) $k_{12} = 0.2$, $k_{23} = 0.2$ and $k_{13} = 0.02$.

(c) $k_{12} = 0.3$, $k_{23} = 0.3$ and $k_{13} = 0.03$, Index = $1..m*n*l$ where $m, n$ and $l$ are the numbers of scan steps in $f_1 = 4700: 50: 5300$, $f_2 = 4700: 50: 5300$ and $f_3 = 4700: 50: 5300$

Fig. 5. Dissipated power of the scanned simulation with

(a) $k_{12} = 0.1, k_{23} = 0.1$ and $k_{13} = 0.01$.

(b) $k_{12} = 0.2, k_{23} = 0.2$ and $k_{13} = 0.02$.

(c) $k_{12} = 0.3, k_{23} = 0.3$ and $k_{13} = 0.03$.

Fig. 6. Axial electric field distribution of different modes:

(a) $2\pi$-mode

(b) $\pi/2$-mode

(c) $\pi$-mode

Fig. 7. Normalized beam-loading conductance as a function of the beam voltage and the phase shift: The color bar represents the value of normalized beam-loading conductance.

Fig. 8. Normalized beam-loading conductance as a function of transit angle between cells and the phase shift: The color bar represent the value of normalized beam-loading conductance

(a) $\beta_e l$, between 1st and 2nd cell

(b) $\beta_e l_{23}$, between 2nd and 3rd cell

Fig. 9. Axial electric field distributions of the designed output cavity system:

(a) $f_1 = 4700\ MHz, f_2 = 5000\ MHz, f_3 = 5250\ MHz$ with $k_{12} = 0.2, k_{23} = 0.2, k_{13} = 0.02$

(b) $f_1 = 4700\ MHz, f_2 = 5050\ MHz, f_3 = 5200\ MHz$ with $k_{12} = 0.2, k_{23} = 0.2, k_{13} = 0.02$

Fig. 10. Comparison of the output power of simulation results and the normalized beam-loading conductance at the fixed beam voltage 68 kV:

(a) $k_{12} = 0.2, k_{23} = 0.2$ and $k_{13} = 0.02$

(b) $k_{12} = 0.2, k_{23} = 0.25$ and $k_{13} = 0.02$

Fig. 11. Reciprocal beam-loading quality factor as a function of the beam voltage:

(a) $f_1 = 4700\ MHz, f_2 = 5050\ MHz, f_3 = 5200\ MHz$ with $k_{12} = 0.2, k_{23} = 0.2, k_{13} = 0.02$

(b) $f_1 = 4700\ MHz, f_2 = 5100\ MHz, f_3 = 5300\ MHz$ with $k_{12} = 0.2, k_{23} = 0.25, k_{13} = 0.02$

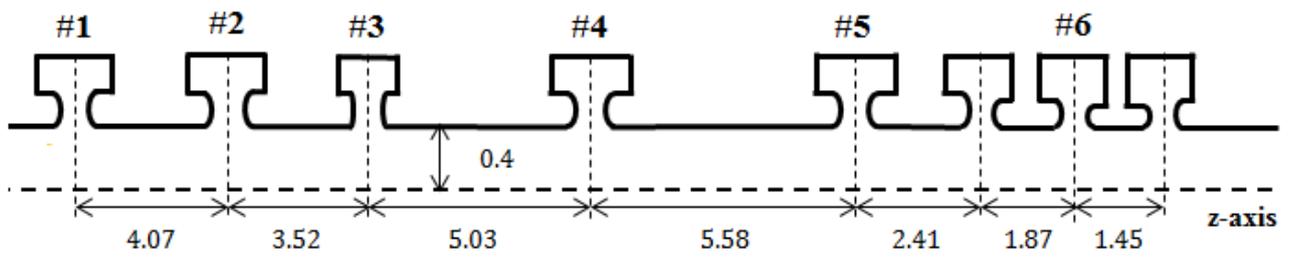

Fig. 1.

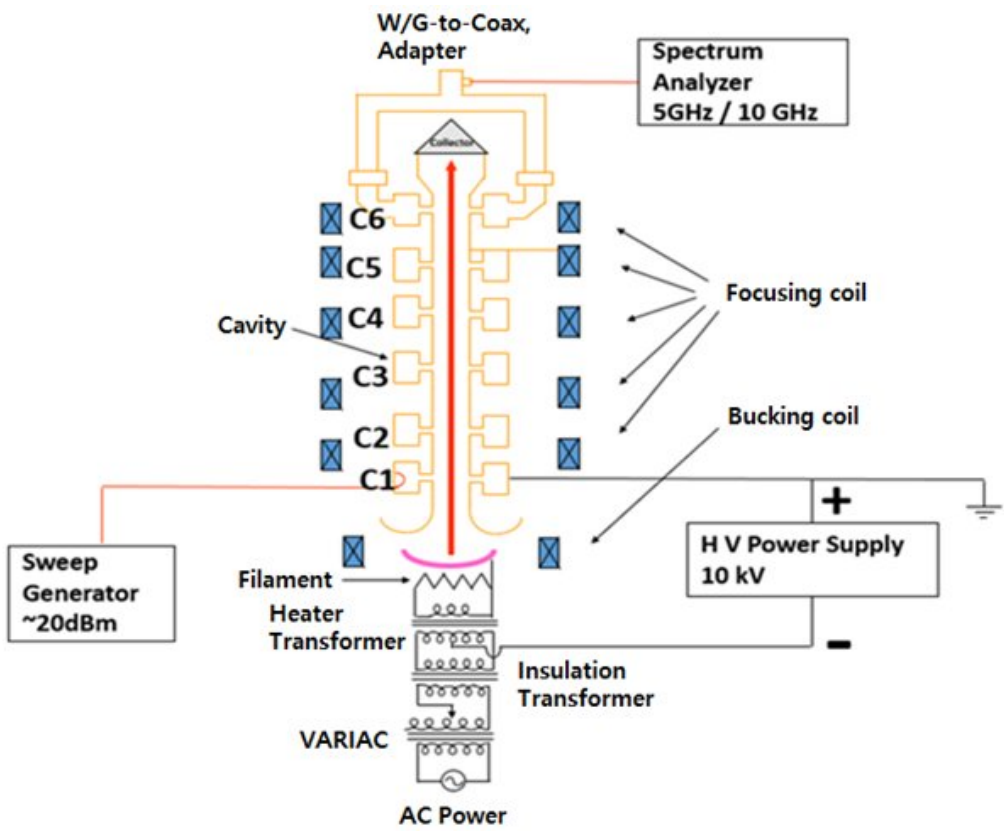

Fig. 2.

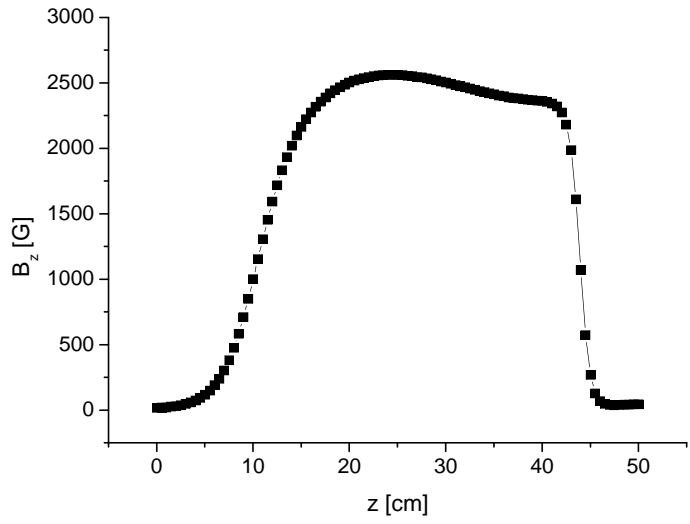

Fig. 3.

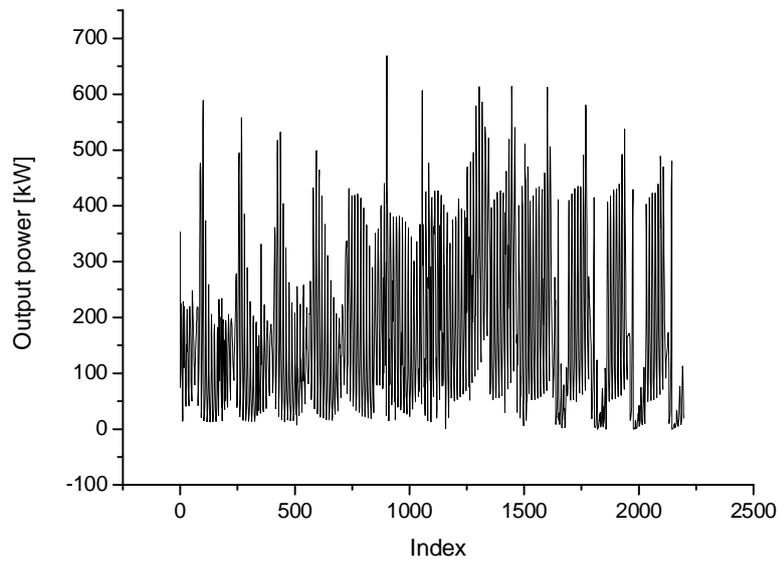

Fig. 4(a).

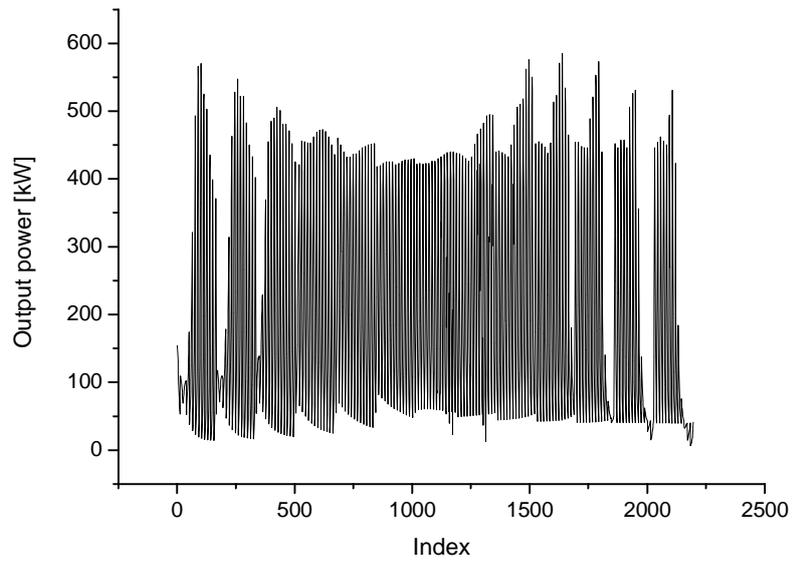

Fig. 4(b).

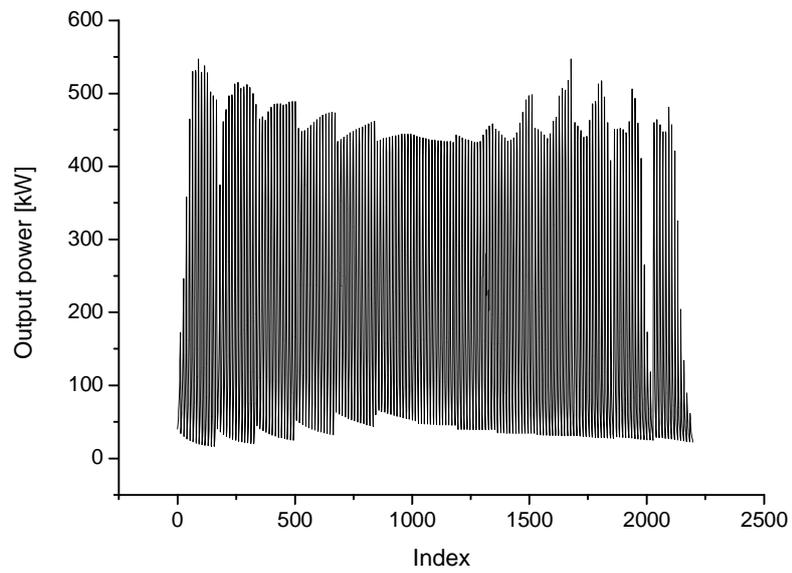

Fig. 4(c).

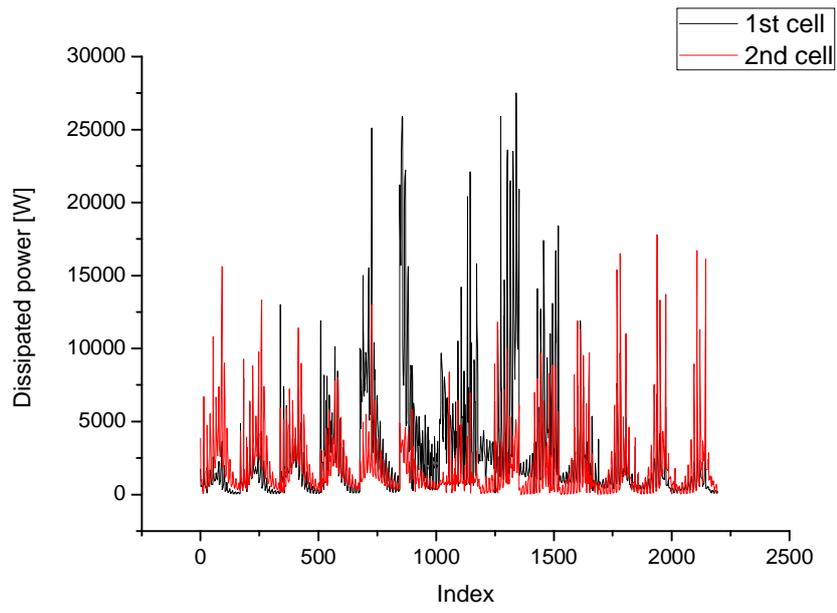

Fig. 5(a).

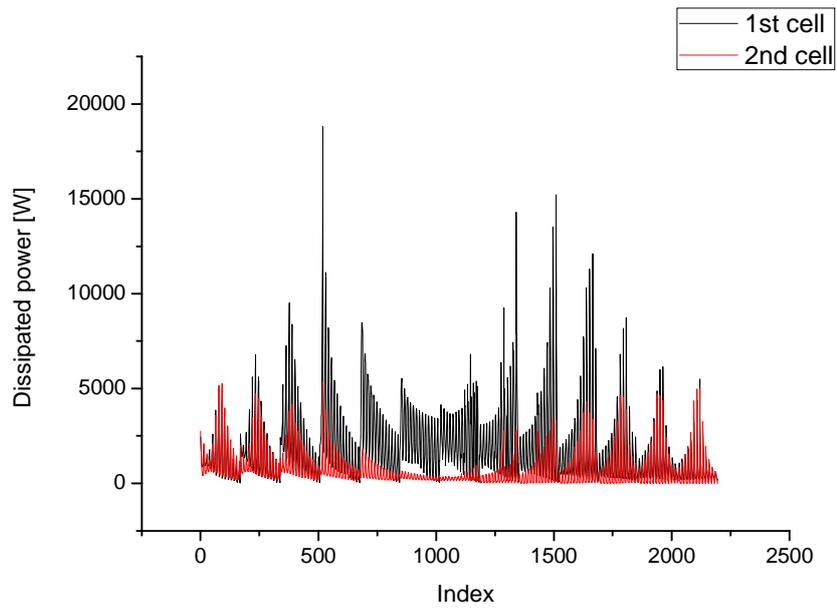

Fig. 5(b).

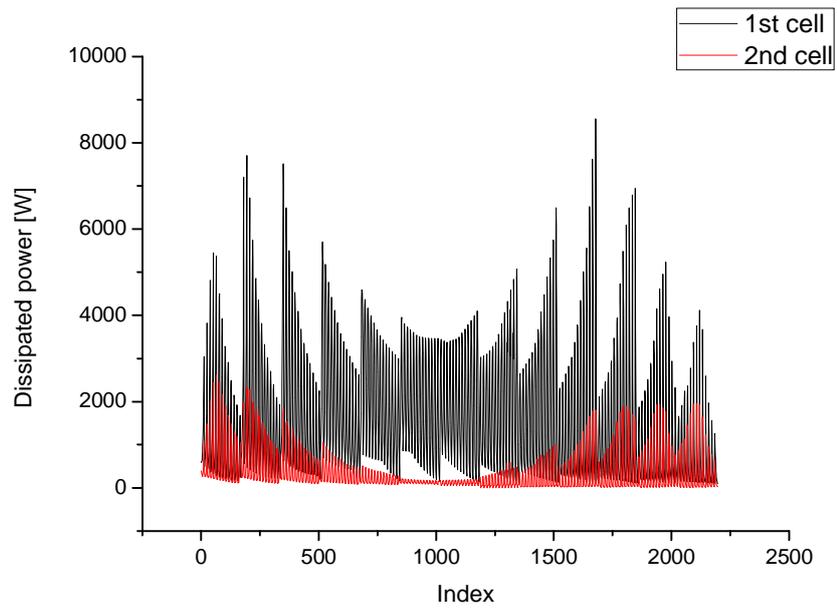

Fig. 5(c).

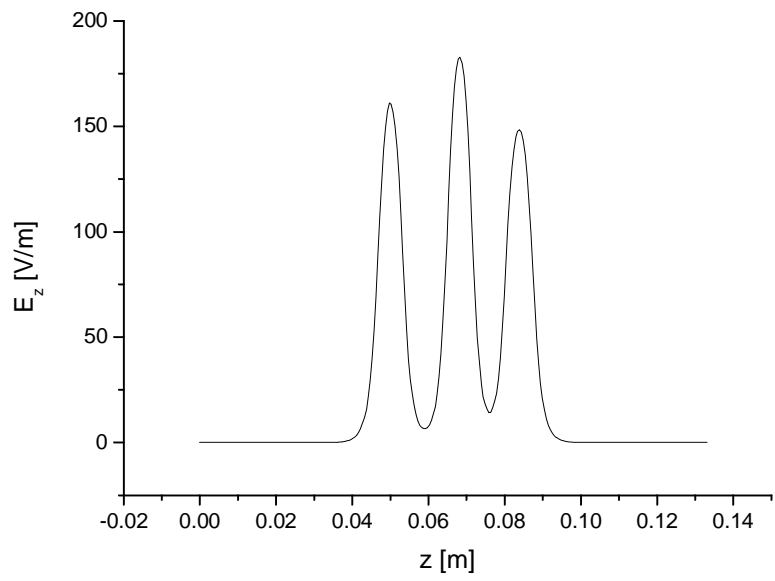

Fig. 6(a).

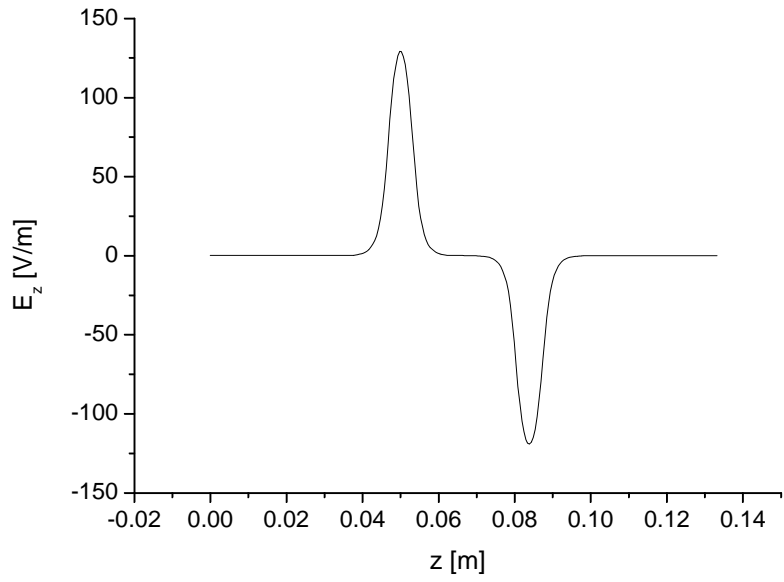

Fig. 6(b).

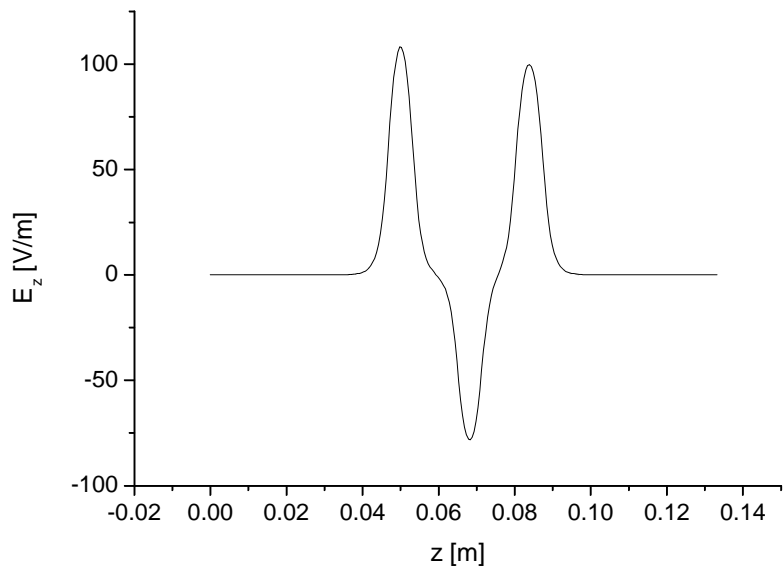

Fig. 6(c).

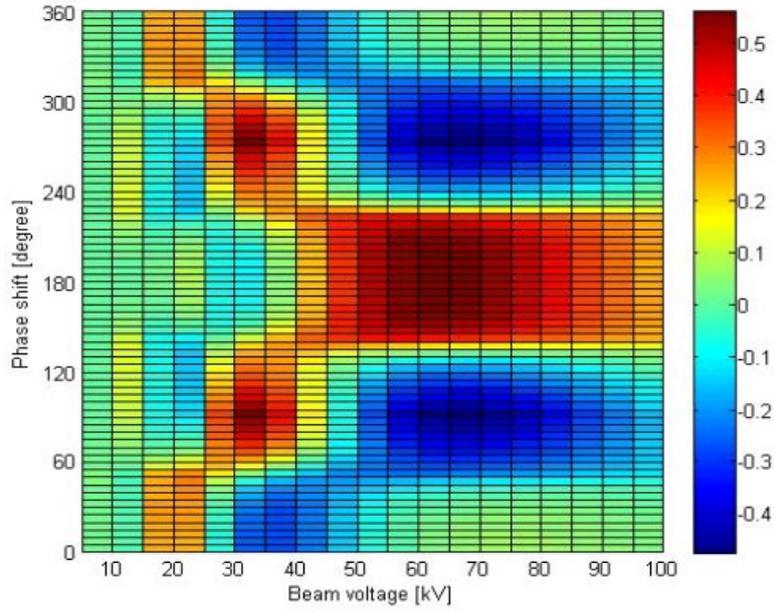

Fig. 7.

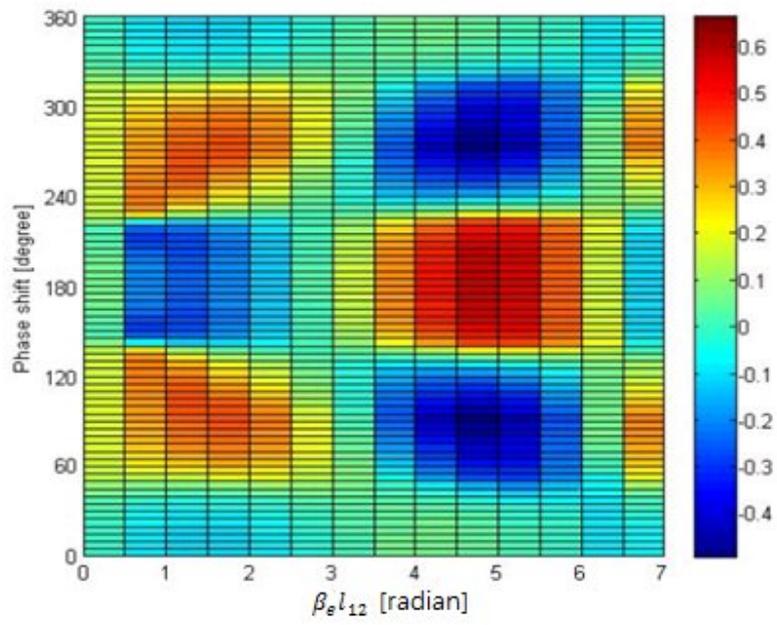

Fig. 8(a).

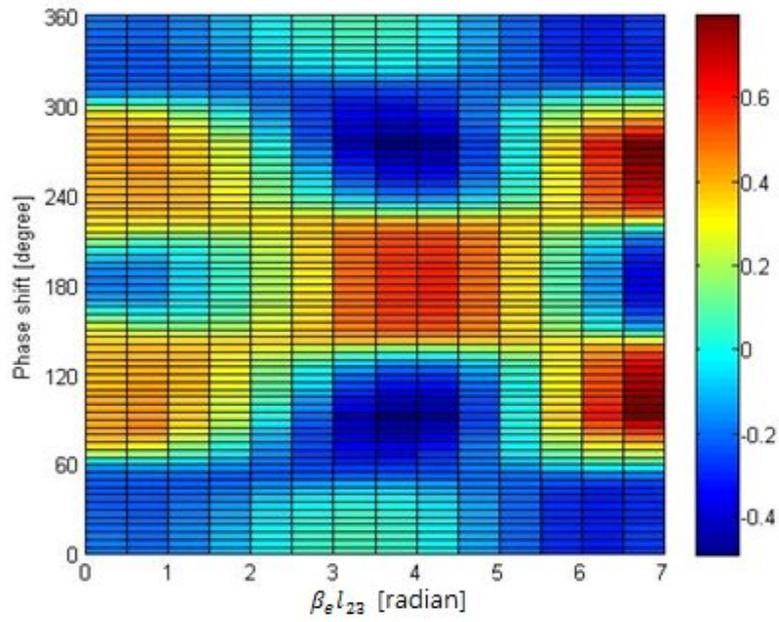

Fig. 8(b).

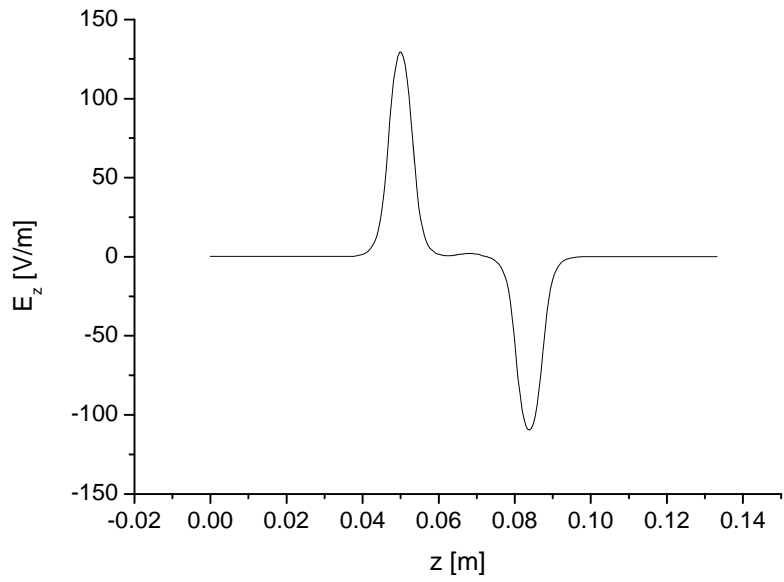

Fig. 9(a).

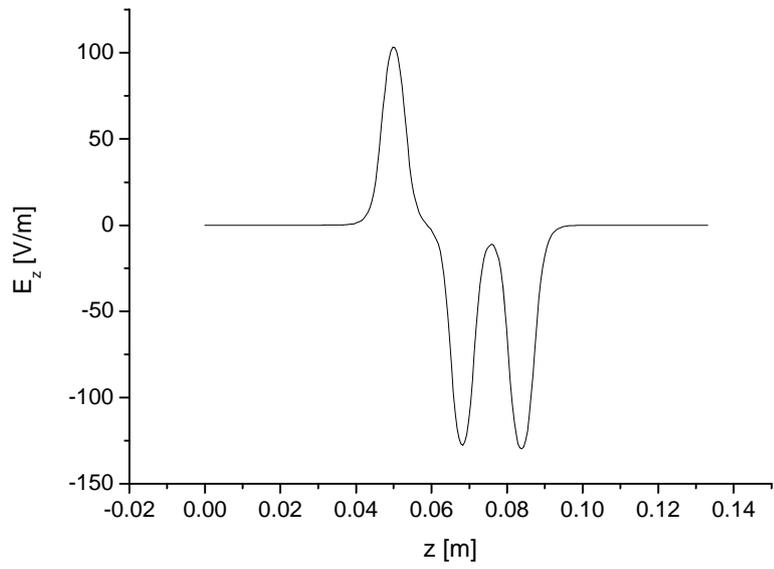

Fig. 9(b).

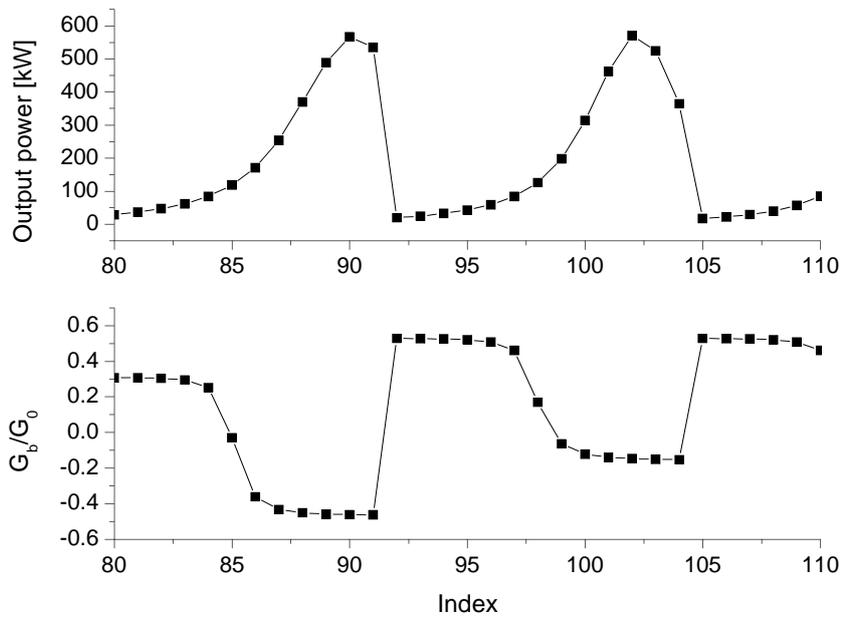

Fig. 10(a).

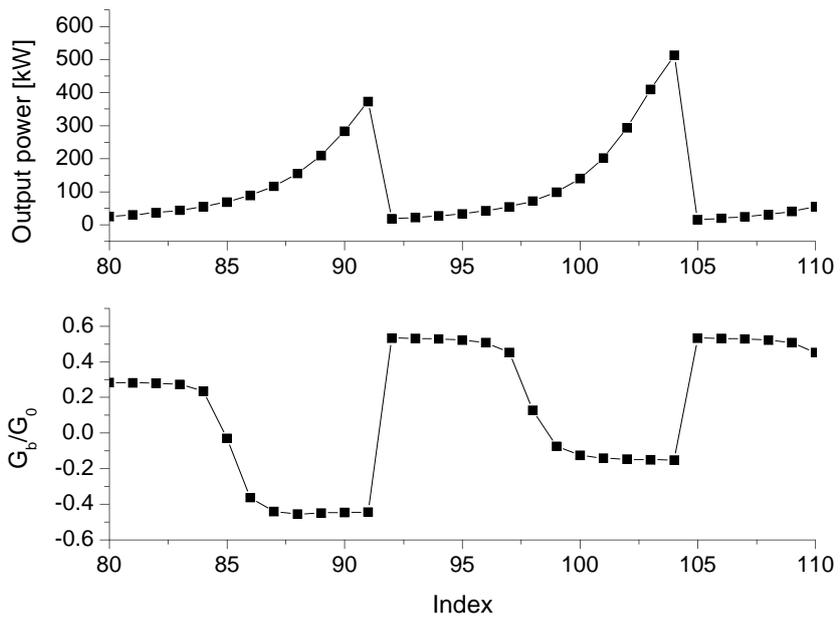

Fig. 10(b).

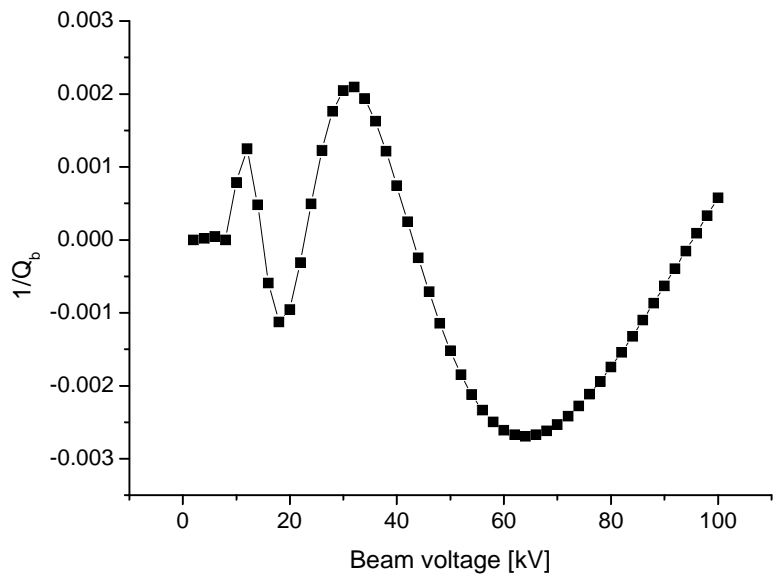

Fig. 11(a).

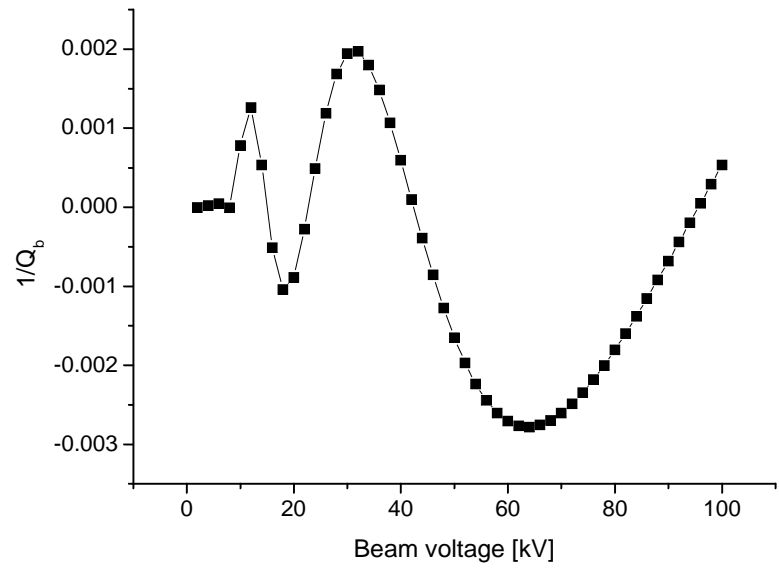

Fig. 11(b).